\documentstyle[dina4,12pt,psfig]{article}
\newcommand{\be}{\begin{equation}}
\newcommand{\ee}{\end{equation}}
\newcommand{\bea}{\begin{eqnarray}}
\newcommand{\eea}{\end{eqnarray}}

\def\p{\partial}

\begin{document}
\title{%
Pairproduction in a Semiclassical Transport Theory%
\thanks{Work supported by BMFT and GSI Darmstadt.}}
\author{%
S. Loh, T.S. Bir\'o and U. Mosel\\
Institut f\"ur Theoretische Physik, Universit\"at Giessen\\
W-35392 Giessen, Germany\\
}
\maketitle
\begin{abstract}
In high-energy nucleon-nucleon-collisions large meson (or $q\bar{q}$)
production rates are observed.
As an alternative to meson exchange mo\-dels here we propose a
nonperturbative treatment
of the quark-antiquark production from the mean field of colliding
nucleon bags.
Source terms for particles and antiparticles are developed in a semiclassical
quark transport description and the time evolution of the quark density
is studied for different mean fields numerically.
\end{abstract}

\section{\bf Introduction}

\bigskip
Transport theories based on the Vlasov equation (BUU/VUU-model
\footnote{Boltzmann-
Uehling-Uhlenbeck or Vlasov-Uehling-Uhlenbeck}) and their relativistic
extensions have been recently used to describe the experimentally
observed scenario in heavy ion collisions. For a recent review and further
references see \cite{buu1}. Motivated by
their success, we have described nucleon-nucleon collisions in the energy
range of a few GeV by
the same equations on the quark level. We have studied especially the dynamical
behaviour of confined quark systems in such reactions \cite{kalm}.
One of the major problems in this so called 'Quark Transport
Theory' is that the production of quark-antiquark-pairs from the mean
field, an important effect due to the low current quark mass, must be
implemented in addition to the Vlasov equation, which describes the
time evolution of the phase space density of a smoothly evolving plasma
semiclassically.\\
The main aim of this work is to study how to treat such pair production
processes within a semiclassical quark transport theory. For this purpose
we first take a look at Klein's Paradox, an analytical description of pair
production in external fields, to get some insight into this process.
Guided by that results we will then propose a practical method how to extend
the usual semiclassical transport model to include these effects.

\section{\bf Klein's Paradox}
First we briefly review the quantum mechanical description of the
scattering of a relativistic spin-1/2- particle on a potential well,
the so called 'Klein's paradox' \cite{klein,dirac}.
Let the incident particle have a momentum
$\vec{p_0} = (p_0,0,0)$ and an energy $E_0 = \sqrt{p_0^2 + m^2}$.
Solving the 1+1-dimensional Dirac equation
\be
  (E_0-V(x))\psi (x) = (-i\hat{\alpha}_x \partial_x + \hat{\beta} m)
  \psi (x)
\ee
with
\be
  V(x) = \left\{ \begin{array}{r@{\quad:\quad}l}
  0 & x<0 \\ V & x \ge 0 \end{array} \right.
\ee
in the two different regions $V(x) = 0$ and $V(x) = V$, we get
the following spinor wave functions for the incoming, reflected and
transmitted particles respectively:
\be
  \psi_{in} = a {1 \choose {p_0 \over E_0 + m}} e^{ip_0x} \; ,
\ee
\be
  \psi_{ref} = c {1 \choose {-p_0 \over E_0 + m}} e^{-ip_0x}
\ee
and
\be
  \psi_{tr} = b {1 \choose {\bar{p}_0 \over E_0 - V + m}}
  e^{i\bar{p}_0x} \; ,
\ee
with
\be
  \bar{p}_0 = \pm \sqrt{(E_0-V)^2 - m^2} \; ,
\ee
where the sign of the momentum $\bar{p}_0$ for the transmitted particle
is not yet fixed. By matching the wave
functions at $x=0$, the unknown coefficients $b$ and $c$ can be determined
and the resulting currents are
\bea
  j_{in} &=& \psi_{in}^{\dagger} \hat{\alpha}_x \psi_{in}
          = a^*a {2p_0 \over E_0 + m} \; , \\
  j_{ref} &=& \psi_{ref}^{\dagger} \hat{\alpha}_x \psi_{ref}
          = -a^*a {(1-\eta^*)(1-\eta) \over (1+\eta^*)(1+\eta)}
          {2p_0 \over E_0 + m} \; , \\
  j_{tr} &=& \psi_{tr}^{\dagger} \hat{\alpha}_x \psi_{tr}
          = -a^*a {4 \over (1+\eta^*)(1+\eta)}
          {2\bar{p}_0 \over V- (E_0 + m)}
\eea
with
\be
  \eta = {\bar{p}_0 \over p_0} {(E_0 + m) \over (E_0 - V + m)} \; .
\ee
Note that the transmitted particles momentum $\bar{p}_0$ is imaginary
in the region $E_0-m<V<E_0+m$ for the potential. If we insist for the
remaining values of the potential that the transmitted current $j_{tr}$
is always positive, then we have to choose
$\bar{p}_0 = + \sqrt{(E_0-V)^2 - m^2}$ if $V < E_0 + m$ and \newline
$\bar{p}_0 = - \sqrt{(E_0-V)^2 - m^2}$ if $V > E_0 + m$. The
reflection and transmission coefficients are then given by
\bea
  \cal{R} &=& {|j_{ref}| \over |j_{in}|} = {(1-\eta)^2 \over (1+\eta)^2}
  \; , \\
  \cal{T} &=& {|j_{tr}| \over |j_{in}|} = {4\eta \over (1+\eta)^2}
\eea
for real values of $\eta$.
Their behaviour as a function of the potential strength can be discussed
as follows:\\
For $V < E_0 - m \;$  $\cal{R}$ grows steadily from 0 to 1 while remaining
equal to unity in the range $E_0 - m < V < E_0 + m$. For $V > E_0 + m$
it drops again and there is
a nonvanishing probability for the particle to penetrate the potential
well. This result is in contradiction to the nonrelativistic treatment,
where a further tunneling for $V>E_0+m$ is forbidden, and was originally
called 'Klein's paradox'.\\
A way out of this puzzle is offered by dropping the single particle
interpretation and analyzing the problem from a many-particle point
of view (hole theory). With the existence of negative energy states
the boundary
condition imposed on the scattering process changes for $V > E_0 + m$.
In this case the lower energy continuum (Dirac sea) on the right overlaps
with the positive energy states on the left (see figure (\ref{see})).
In this case the particles in the negative energy states on the right
could leave the sea to the left, thus creating holes, i.e. antiparticles
travelling to the right. This changes the sign of the transmitted
particles' momentum
\be
  \bar{p}_0 = +\sqrt{(E_0-V)^2-m^2)} \; .
\ee
The result is, that $\cal{R}$ becomes larger than 1, which means that
the reflected current is larger than the incoming one and the transmitted
current is negative, though the continuity equation
\be
  j_{in}+j_{ref} = j_{tr}
\ee
is nevertheless fulfilled. This continuity equation now describes
charge conservation and the
currents $j_{in}$, $j_{ref}$ and $j_{tr}$ are charge-current densities.
The negative value of $j_{tr}$ is therefore interpreted as an outgoing
current of particles of opposite charge, namely antiparticles.
In this sense we are dealing with a pair production process, which is
illustrated in figure (\ref{see}).\\
For a later comparison to transporttheoretical calculations we investigated
also the scattering of Gaussian wave packets
\be
  \psi (x) = N \int {dp \over 2\pi} {m \over E} e^{-p^2 \Delta^2 /2}
             u(p) e^{ipx} \; ,
\ee
with a width $\Delta$, normalization constant $N$ and positive
energy spinor $u(p)$.
The resulting reflection coefficient $\cal{R}$
for both the single-particle and the particle-hole interpretation
are shown in figure (\ref{refl}), where, for we show a very broad
wave packet, almost no deviations from the plane wave case can be seen.

\section{\bf Continous potentials}

We now want to study how the pair production process is affected if
the edge of the potential well is smoothed out. For this purpose
we take a Woods-Saxon-potential,
\be
  V(x) = V \big[ {1 \over 1+\exp(-ax)} - {1 \over 2} \big] \; ,
\ee
where $a$ is the diffuseness parameter and $V$ is the strength of
the potential.
Solving the Dirac equation with the same initial condition as
above, we get \cite{sauter}
\bea
  \cal{R} &=& \frac{\cosh{({2\pi \over a}V)} - \cosh{({2\pi \over a}
  (p_0+\bar{p}_0))}} {\cosh{({2\pi \over a}V)} - \cosh{({2\pi \over a}
  (p_0-\bar{p}_0))}} \; , \\*[0.3cm]
  \cal{T} &=& \frac{\cosh{({2\pi \over a}(p_0-\bar{p}_0))} - \cosh{({2\pi \over
a}
  (p_0+\bar{p}_0))}} {\cosh{({2\pi \over a}V)} - \cosh{({2\pi \over a}
  (p_0-\bar{p}_0))}} \; ,
\eea
for the reflection and transmission coefficient respectively within
the single particle interpretation and \cite{dipl}
\bea
  \cal{R} &=& \frac{\cosh{({2\pi \over a}V)} - \cosh{({2\pi \over a}
  (p_0-\bar{p}_0))}} {\cosh{({2\pi \over a}V)} - \cosh{({2\pi \over a}
  (p_0+\bar{p}_0))}} \; , \\*[0.3cm] \label{transh}
  \cal{T} &=& \frac{\cosh{({2\pi \over a}(p_0+\bar{p}_0))} - \cosh{({2\pi \over
a}
  (p_0-\bar{p}_0))}} {\cosh{({2\pi \over a}V)} - \cosh{({2\pi \over a}
  (p_0+\bar{p}_0))}} \; ,
\eea
in the particle-hole picture. In the latter case
$\bar{p}_0$ has been changed to $-\bar{p_0}$ due to the different boundary
conditions.\\
Consider now the transmission coefficient
$\cal{T}$ (\ref{transh}) for large values of $V$ and $E_0=V/2$, in which case
$p_0 = \bar{p}_0$. Then $\cal{T}$ converges
to
\be
  {\cal{T}} \to \exp{(- \tilde{c} {\lambda \over \lambda_c})} \; ,
\ee
where $\lambda = 1/a$ is the surface thickness of the potential,
$\lambda_c = \hbar / mc$
the Compton wave length and
\be
  \tilde{c} = 4\pi \sqrt{{1-\beta} \over {1+\beta}}
\ee
is proportional to the relativistic Doppler-factor. This
means that the decisive parameter for the pair production process is the
ratio of the surface thickness of the potential to the Compton
wave length of the particle. This expectation is borne out in
figure (\ref{reflam}), which shows that $\cal{T}$ is significantly
different from zero only for $\lambda < \lambda_c$.\\
On the other hand, if we are dealing with quark systems, usually
$\lambda \ll \lambda_c$ is expected due to the low current quark mass,
causing considerable pair production.

\section{\bf The quark transport model}

We have seen in the last section, that the $q$-$\bar{q}$ production
becomes important when the potentials seen by the quarks change
significantly over the distance of the quark's Compton wavelength.
This fact seems to present a major obstacle to any attempt to incorporate
the mean-field $q$-$\bar{q}$ production into a semiclassical
transporttheoretical description of nucleon-nucleon collisions, which
can be expected to describe only phenomena connected with slow changes
of the potentials. In this section we therefore develop a practical
method how to overcome this difficulty by adding explicit source-terms
to the relativistic Vlasov equation.\\
The basic ingredient of any relativistic transport theory is the Wigner
function
\be
  W(x,p) = \int dR \; \bar{\psi}(x-{R \over 2}) \psi (x+{R \over 2})
           \exp^{-ipR} \; ,
\ee
the quantum mechanical analog to the classical phase space density,
which is used in deriving the equation of motion.
In the 1+1-dimensional case we use the following ($2\times 2$)
Dirac matrices
\bea
  \Gamma_1 = \left( \begin{array}{cc} 1 & 0 \\ 0 & 1 \end{array} \right)
  \; &,& \;
  \Gamma_2 = \left( \begin{array}{cc} 1 & 0 \\ 0 & -1 \end{array} \right)
  \\
  \Gamma_3 = \left( \begin{array}{cc} 0 & 1 \\ -1 & 0 \end{array} \right)
  \; &,& \;
  \Gamma_4 = \left( \begin{array}{cc} 0 & 1 \\ 1 & 0 \end{array} \right)
  \; .
\eea
{}From the Dirac equation we obtain a Vlasov equation in the standard
way \cite{egvhsg,remler}
\be\label{vlas}
  tr(DW) = 0 \; ,
\ee
where
\be
  D = \Gamma_2 \p_t + \Gamma_3 \p_x + \tilde{F}\Gamma_2 \p_p
        - \tilde{F}\Gamma_3 \p_E
\ee
is a matrix-valued differential operator and the force term
is $\tilde{F} = \p_x V(x)$ in our special case of a static potential
$A_{\mu} =(V(x),0)$. Equation (\ref{vlas}) can be rewritten in
terms of a scalar phase space distribution function
$F(x,\Pi) = {1 \over 2} tr(W)$, using the canonical momentum
$\Pi_{\mu} = p_{\mu} - eA_{\mu}$, as
\be\label{vlas2}
  (\Pi_{\mu} \p^{\mu} + \tilde{F}_{\mu\nu}\Pi^{\nu} \p^{\mu}_{\Pi})
  F(x,\Pi) = 0 \; .
\ee
Using furthermore the mass shell constraint $F(x,\Pi) = \delta ( \Pi_{\mu}
\Pi^{\mu} - m^2) f(x,\Pi_x)$ we end up with the familiar Vlasov
equation for the phase space density $f(x,p,t)$,
\be\label{vlas3}
  (\p_t + {p \over E} \p_x + \tilde{F} \p_p) f(x,p,t) = 0 \; ,
\ee
since $\Pi_x = p$.\\
As shown in \cite{dipl}, the Vlasov equation (\ref{vlas2}) is derived in
leading order in a semiclassical expansion. It has to be modified in
order to include pair production.
This we illustrate with the example of a static external potential
well. An extension to more general potentials will be discussed later
on.\\
In analogy to the separation of the quantum mechanical problem into
three wave functions we split up the semiclassical phase space density
into three parts, too:
\be
  f(x,p,t) = f_{in}(x,p,t) + f_{ref}(x,p,t) + f_{tr}(x,p,t) \; .
\ee
In the following we first omit the Vlasov force term $\tilde{F}$ of
equation (\ref{vlas3}) since it presents no difficulties and stems from
a semiclassical derivation.
We then look for a set of differential equations
\be
  D f_j = (\p_t + {p \over E} \p_x) f_j = S_j  \; ,
\ee
where the index $j$ means the incoming, reflected and transmitted part
respectively,
$j \in \{in,ref,tr\}$ and $D = \p_t + p/E \p_x$ is the Vlasov
differential operator. The $S_j$ are source terms for the phase space
densities, which si\-mu\-late the correct quantum mechanical
pair production process.\\
We generate the dynamics now exclusively by the source terms $S_j$. In a
dynamical simulation the $S_j$ have to 'absorb' $f_{in}$ and 'emit'
$f_{ref}$ and $f_{tr}$. In addition, they are subject to the condition
of charge conservation.
An ansatz fulfilling both conditions has the form
\bea
  \label{source1}
  (\p_t + {p \over E}\p_x)  f_{in} (x,p,t) &=&- \delta (x) {p \over E}
  f_{in} (x,p,t) \\*[0.3cm]
  \label{source2}
  (\p_t - {p \over E}\p_x) f_{ref} (x,-p,t) &=& \delta (x) {p \over E}
  f_{in} (x,p,t) {\cal{R}} (p,V)\\*[0.3cm]
  \label{source3}
  (\p_t - {\bar{p} \over \bar{E}}\p_x) f_{tr} (x,p,t) &=& - \delta (x)
  {\bar{p} \over \bar{E}} f_{in} (x,p,t) {\cal{T}} (p,V) \\*[0.3cm]
  \mbox{with} \quad E = \sqrt{p^2+m^2} \quad & & \mbox{and} \quad
  \bar{E} = \sqrt{\bar{p}^2+m^2} \; .
\eea
In order to make a numerical study feasible, we solve equations
(\ref{source1}) - (\ref{source3}) using a 'weighted test
particle method' \cite{tohyama,dipl}, where the phase space density is
sampled by an ensemble of 'testparticles',
\bea
  \label{testp1}
  f_{in} (x,p,t) &=& 2\pi \sum_{i=1}^N w_i^{in}(t) \delta(x-x_i^{in}(t))
  \delta(p-p_i^{in}(t)) \; , \\
  \label{testp2}
  f_{ref} (x,p,t) &=& 2\pi \sum_{i=1}^N w_i^{ref}(t) \delta(x-x_i^{ref}(t))
  \delta(p-p_i^{ref}(t)) \\ \mbox{and} \quad
  \label{testp3}
  f_{tr} (x,p,t) &=& 2\pi \sum_{i=1}^N w_i^{tr}(t) \delta(x-x_i^{tr}(t))
  \delta(p-p_i^{tr}(t)) \; ,
\eea
with time dependent weight factors $w_i^j (t)$. Inserting eq.
(\ref{testp1}) - (\ref{testp3}) into (\ref{source1}) - (\ref{source3})
we get the familiar Hamiltonian equations of motion for the coordinates
and the momenta of the testparticles
\be
  \dot{x}_i = {p_i \over E_i} \quad , \quad
  \dot{p}_i = -\p_x V
\ee
and the following equations for the weight factors:
\bea\label{weight1}
  {d \over dt} w_i^{in}(t) &=& - \delta (x_i^{in}(t)) {p_i^{in} \over
  E_{p_i^{in}}} w_i^{in}(t) \; , \\
  {d \over dt} w_i^{ref}(t) &=& \delta (x_i^{in}(t)) {p_i^{in} \over
  E_{p_i^{in}}} w_i^{in}(t) {\cal{R}} (p_i^{in},V) \\
  \label{weight3} \mbox{and} \quad
  {d \over dt} w_i^{tr}(t) &=& - \delta (x_i^{in}(t)) {\bar{p}_i^{in}
  \over E_{\bar{p}_i^{in}}} w_i^{in}(t) {\cal{T}} (p_i^{in},V) \; .
\eea
The solution of (\ref{weight1}) - (\ref{weight3}) for the potential step
at $x=0$ is simply given by
\bea
  \label{weight4}
  w_i^{in} (t) &=& \theta ( -x_i^{in}(t)) \; , \\
  \label{weight5}
  w_i^{ref}(t) &=& {\cal{R}} (p_i^{in},V) (1-w_i^{in}(t)) \\
  \label{weight6} \mbox{and} \quad
  w_i^{tr}(t) &=& {\cal{T}} (p_i^{in},V) (1-w_i^{in}(t))
  {\bar{p}_i^{in} \over E_{p_i^{in}}} {E_{p_i^{in}} \over p_i^{in}} \;
  .
\eea
We can now clearly see the effect of the source terms $S_j$. If we
take
\be
  w_i^{in}(0) = 1 \quad , \quad w_i^{ref,tr}(0) = 0
\ee
as an initial condition,
we describe an incoming current from the left, with the weight
factors of the testparticles vanishing as they reach the
potential well.
Simultaneously, the corresponding reflected and transmitted
test(anti)particle weights are generated with factors
\be
{\cal{R}} \quad \mbox{and}  \quad
{\bar{p}_i^{in} \over E_{\bar{p}_i^{in}}} {E_{p_i^{in}} \over p_i^{in}}
\cal{T} \; .
\ee
\\
In order to demonstrate the feasibility to include these phenomena into
a time-dependent transport theory with moving wave packets we show in
figure (\ref{dens}) a typical time evolution of the charge
density $\rho (x,t) = \int {dp \over 2\pi} f(x,p,t) $ obtained in the
transport model. It can be seen that the density increases strongly
when the potential well is reached. Particles and
antiparticles are emitted in opposite directions, respectively.\\
We also calculated the asymptotic values of the currents,
\be
j_{\mu} = \int tr( \Gamma_{\mu} W) {dp\over 2\pi} \; ,
\ee
for time dependent wave packets.
Then we compared the numerically obtained reflection coefficients with
the analytical ones. The result of this comparison
is shown in figure (\ref{refanti}).
It can be seen that in the special case of a sharp potential well
the transport model describes the pair production via the source
terms eqs. (\ref{source1}) - (\ref{source3}) with a high accuracy.\\
Figure (\ref{nqqbar}) presents the results of studying Woods-Saxon shape
potentials with different surface thickness. In order to gain some qualitative
insight into the pair production mechanism we used a potential height of
$V = 0.3 \, GeV$ and a bare quark mass of $m = 0.01 \, GeV$. The rising length
varied between $0.025 \,\lambda_c$ and $0.25 \,\lambda_c$, where $\lambda_c$ is
the Compton wavelength of a quark. The different lines correspond to
stationary plane waves for the quarks with energies ranging from
$E = \, 0.1 GeV$ to $E = 0.3 \, GeV$.
The full circles represent the results of the transport model using a
Gaussian wave packet around an energy of $E = 0.1 \,GeV$ and having a width
of $\Delta = 1 \,fm$. The stars stand for a similar calculation using
$E = 0.26 \,GeV$ and the same width. One observes the general tendency of
Gaussian wave packets producing more quark antiquark pairs on steep
potentials than the corresponding plane waves. For potentials having a
rising length bigger then $\lambda = 0.08 \, \lambda_c$ the Gaussian wave
packets are less effective. This deviation is, however, suppressed at
lower energies.\\
Finally figure (\ref{pprate}) shows the energy dependence of the pair
production
rate from Gaussian wave packets for different Woods-Saxon potentials
($\lambda = 0.5 \, fm$ to $\lambda = \, 5 fm$) obtained in the transport model.
The stars belong to a rising length of $\lambda = 0.5 \, fm$, the squares to
$\lambda = 1 \, fm$, the triangles to $\lambda = 2 \, fm$ and the circles to
$\lambda = 5 \, fm$. For comparison the corresponding plane wave results
are shown by the solid, dashed and dotted lines. The general suppression
tendency of this pair production mechanism at energies approaching the
potential barrier is clearly seen both for plane waves and for wave
packets. Though the wave packets have a large spread in energy the
pair production rate is rather constant over a wide energy regime.
In this case even at intermediate
energy values in the order of the potential height a nonvanishing
pair production rate can be observed.\\
Summarizing we developed a semiclassical transport model description
of quark antiquark pair production on external potential gradients, which
we tested by comparing results using Gaussian wave packets and plane waves.
The source term prescription eq. (\ref{source1}) - (\ref{source3})
proposed in this paper gives numerically accurate results and opens up a
perspective of more realistic simulations of complex dynamical quark
systems.

\newpage
\begin{figure}
\centerline{\psfig{figure=c:/users/stefanl/tex/pics/see.ps,width=15cm}}
\caption{Energy levels of the Dirac equation in an external field
$(V>E+m)$.}
\label{see}
\end{figure}
\begin{figure}
\centerline{\psfig{figure=c:/users/stefanl/tex/pics/ratvonv.ps,width=15cm}}
\caption{Reflection coefficient ${\cal{R}}_s$ for a stationary (solid line)
and ${\cal{R}}_t$ for a time dependent (stars)
hole theoreric calculation. The parameters $m=5 \, fm^{-1}$, $E=9.5 \,
fm^{-1}$, $\Delta=10 \, fm$ were used.}
\label{refl}
\end{figure}
\begin{figure}
\centerline{\psfig{figure=c:/users/stefanl/tex/pics/rvonlam.ps,width=15cm}}
\caption{Reflection and transmission coefficients ${\cal{R}}_s$, ${\cal{T}}_s$
and ${\cal{R}}_t$, ${\cal{T}}_t$ as functions of the
length of rise of the potential $\lambda$. The parameter set $m=5 \, fm^{-1}$,
$E=9.5 \, fm^{-1}$, $\Delta=10 \, fm$ is used.}
\label{reflam}
\end{figure}
\begin{figure}
\centerline{\psfig{figure=c:/users/stefanl/tex/pics/rhocxt.ps,width=15cm}}
\caption{Time evolution of the charge-density $\rho_B$ within the transport
model using an energy of $E = 9.5 \, fm^{-1}$, quark mass $m = 5 \, fm^{-1}$,
wave packet width $\Delta = 1.5 \, fm$ and potential strength
$V = 20 \, fm^{-1}$.}
\label{dens}
\end{figure}
\begin{figure}
\centerline{\psfig{figure=c:/users/stefanl/tex/pics/refvlas.ps,width=15cm}}
\caption{Reflection coefficients for the plane wave (${\cal{R}}_s$),
for the wave packet (${\cal{R}}_t \, Dirac$) and for the transport
model (${\cal{R}}_t \, Vlasov$) calculation.
The parameters of the investigated wave packets are as follows:
$m=5 \, fm^{-1}$, $E=9.5 \, fm^{-1}$, $\Delta=10 \, fm$.}
\label{refanti}
\end{figure}
\begin{figure}
\centerline{\psfig{figure=c:/users/stefanl/tex/pics/nqqbar.ps,width=15cm}}
\caption{Number of produced quark antiquark pairs for different energies
$E$ in the range of $0.1$ to $0.3 \, GeV$. The full dots represent the
results of the transport model using Gaussian wave packets with the energy
$E = 0.1 \, GeV$, while the stars stand for $E = 0.26 \, GeV$.
The remaining parameters are
$m = 0.01 \, GeV$, $\Delta = 1.0 \, fm$, $V = 0.3 \, GeV$.}
\label{nqqbar}
\end{figure}
\begin{figure}
\centerline{\psfig{figure=c:/users/stefanl/tex/pics/pprate.ps,width=15cm}}
\caption{Pair production rate in an external Woods-Saxon potential for
different stiffness parameters.
The stars belong to a transport model calculation with Gausian wave
packets and a rising length of $\lambda = 0.5 \, fm$, the squares to
$\lambda = 1 \, fm$, the triangles to $\lambda = 2 \, fm$ and the circles to
$\lambda = 5 \, fm$. The remaining parameters are $m = 0.01 \, GeV$,
$\Delta = 1.0 \, fm$ and $V = 0.3 \, GeV$.}
\label{pprate}
\end{figure}

\begin{thebibliography}{99}
%
\bibitem{buu1} B. Blttel, V. Koch and U. Mosel,
Rep. Prog. Phys. {\bf 56} (1993) 1
%
\bibitem{kalm} U. Kalmbach, T. Vetter, T. S. Bir\'o, and U. Mosel,
{\em A Quark Transprot Theory to describe Nucleon-Nucleon Collisions},
Nucl. Phys. A, in press
%
\bibitem{klein} O. Klein, Z. f. Physik {\bf 53}, (1929), 157
%
\bibitem{dirac} P.A.M. Dirac, Proc. Roy. Soc. {\bf 117}, (1928), 612
%
\bibitem{sauter} F. Sauter, Z. f. Physik {\bf 73}, (1931), 547
%
\bibitem{dipl} S. Loh, {\em Paarproduktion in einer
Transporttheorie}, Diplomathesis, Giessen (1992)
%
\bibitem{egvhsg} H. Th. Elze, M. Gyulassy, D. Vasak, H. Heinz, H. Stcker,
W. Greiner, Mod. Phys. Lett. A2, 451, (1987)
%
\bibitem{remler} E. A. Remler, Phys. Rev. D, {\bf 16}, (1977), 3464
%
\bibitem{tohyama} M. Tohyama, E. Suraud, {\em Weighted Particle Method for
Solving the Boltzmann Equation}, (GANIL preprint) (1988)

\end{thebibliography}
\end{document}